\documentclass[twocolumn,aps,prb,showpacs,superscriptaddress]{revtex4}
\usepackage[dvips]{graphicx}
\usepackage{bm}
\begin{document}
\title{Josephson (001) tilt  grain boundary junctions of high temperature
superconductors}
\author{Gerald B. Arnold}
\email{garnold@nd.edu} \affiliation{Department of Physics,
University of Notre Dame, Notre Dame, IN}
\author{Richard A. Klemm}
\email{klemm@phys.ksu.edu} \affiliation{Department of Physics,
Kansas State University, Manhattan, KS 66506}
\date{\today}
\begin{abstract}
We calculate the Josephson critical current  $I_c$ across in-plane
(001) tilt grain boundary junctions of high temperature
superconductors.
 We
solve for the  electronic states corresponding to the
electron-doped cuprates, two slightly different hole-doped
cuprates, and an extremely underdoped hole-doped cuprate in each
half-space, and weakly connect the two half-spaces by either
specular or random Josephson tunneling. We treat  symmetric,
straight,  and fully asymmetric junctions with  $s$-,
extended-$s$, or $d_{x^2-y^2}$-wave order parameters. For
symmetric junctions with random grain boundary tunneling, our
results are generally in agreement with the Sigrist-Rice form for
ideal junctions that has been used to interpret
``phase-sensitive'' experiments consisting of such in-plane grain
boundary junctions. For specular grain boundary tunneling across
symmetric junctions, our results depend upon the Fermi surface
topology, but are usually rather consistent with the random facet
model of Tsuei {\it et al.} [Phys. Rev. Lett. {\bf 73}, 593
(1994)].  Our results for asymmetric junctions of electron-doped
cuprates  are in agreement with the Sigrist-Rice form. However,
our results for asymmetric junctions of hole-doped cuprates show
that the details of the Fermi surface topology and of the
tunneling processes are both very important, so that the
``phase-sensitive'' experiments based upon in-plane Josephson
junctions are less definitive than has generally been thought.
\end{abstract}
 \vskip0pt\vskip0pt \maketitle

\section{Introduction}

Because of its relation to  the mechanism for superconductivity in
the high temperature superconducting compounds (HTSC), there has
long been a raging debate regarding their orbital symmetry of the
superconducting order parameter
(OP).\cite{TKreview,Mannhartreview,Mueller,Klemmreview}  Only
``phase-sensitive'' experiments involving Josephson tunneling can
distinguish the OP from the non-superconducting
pseudogap.\cite{TKreview,Klemmreview,pseudogap}  Although the
first in-plane phase-sensitive experiment on
YBa$_2$Cu$_3$O$_{7-\delta}$ (YBCO) suggested a dominant $s$-wave
OP,\cite{ChaudhariLin} experiments using tricrystal films of YBCO,
Bi$_2$Sr$_2$CaCu$_2$O$_{8+\delta}$ (Bi2212), and
Nd$_{1.85}$Ce$_{0.15}$CuO$_{4+\delta}$ (NCCO),
 and  tetracrystal films of YBCO and La$_{2-x}$Ce$_x$CuO$_{4-y}$
 (LCCO)
suggested that these materials had a dominant $d_{x^2-y^2}$-wave
OP.\cite{TKreview,TKYBCO,TKBi2212,TKhole,TKNCCO,MannhartSQUID,Chesca}

Very different results were obtained for $c$-axis junctions. Low
temperature $T$ $c$-axis  Josephson junctions between
 Pb and  YBCO, Bi2212, and NCCO  suggested varying
amounts of an $s$-wave OP
component.\cite{Dynes,Clarke,Katz,KleinerMoessle,Woods}  More
recently, three Bi2212 $c$-axis twist Josephson junction
experiments showed that the OP has at least a substantial, and
possibly a dominant $s$-wave component for $T$ up to the
transition temperature
$T_c$.\cite{Klemmreview,Li,Takano,Klemm,Latyshev} Here we search
for a reason for these qualitative differences.\cite{Klemmreview}

To date, most theoretical treatments of in-plane (001) tilt grain
boundary (GB) junctions either used  the   Ginzburg-Landau (GL)
approach,\cite{SigristRice} or  assumed  a circular in-plane Fermi
surface (FS)  cross-section, and that  the FS-restricted
$d_{x^2-y^2}$-wave OP $\propto\cos(2\phi_{\bm
k})$.\cite{TanakaKashiwaya,Asano,Barash,Kalenkov,Ilichev} Most of
those treatments focussed upon the single quasiparticle density of
states on the surface, rather than the critical current across the
misalignment grain boundary junctions.  The only previous
treatment that included a tight-binding FS was that of Shirai {\it
et al.}, for which the FS consisted of  very small pockets
centered at the corners of the first Brillouin zone
(BZ).\cite{Shirai}  Their chosen FS was extremely different from
any that have been extracted from angle-resolved photoemission
spectroscopy (ARPES) experiments.\cite{bks} Shirai {\it et al.}
studied both the single quasiparticle and supercurrent properties
of  GB junctions constructed from four specific
surfaces.\cite{Shirai} Here we explicitly take account of the
tight-binding hole and electron-doped FS's observed using ARPES
and of the surface boundary conditions (BC's) at the interfaces.

We find that a $d_{x^2-y^2}$-wave OP can be consistent with the
tricrystal experiments for electron-doped cuprates, for which the
two-dimensional  (2D) FS cross-section is nearly isotropic. For
hole-doped cuprates, the situation is found to be inconclusive, as
one FS cross-section is consistent with the tricrystal experiments
for random GB tunneling, but not for specular GB tunneling.
Another, slightly different hole-doped FS, on the other hand, is
consistent with the tricrystal experiments for specular tunneling,
but not for random tunneling.
 The FS of Shirai {\it et al.} is inconsistent with most experiments
 on the hole-doped cuprates, including the tricrystal experiments, for both specular and random GB tunneling.
These combined results suggest
 that defects and  small changes in the FS topology play essential roles in interpreting the
in-plane GB junction experimental results.\cite{TKreview} An
important modification to the tetracrystal experiment is also
warranted.\cite{Mannhartreview}

\begin{figure}\includegraphics[width=0.45\textwidth]{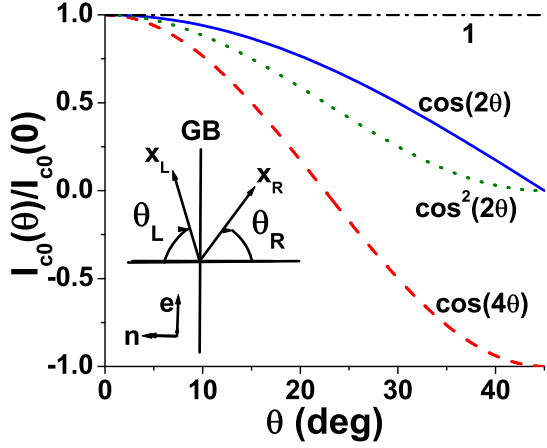}
\caption{(Color online) Plots of $I_{c0}(\theta)/I_{c0}(0)$  for a
$d_{x^2-y^2}$-wave OP with straight (dotted green), symmetric
(dotted green) and asymmetric (solid blue) GB junctions  in the
ideal GL model, and straight (dot-dashed black), symmetric (dashed
red) and asymmetric (solid blue) GB junctions in the faceted GL
model. Inset: Sketch of a (001) tilt GB junction.}\label{fig1}
\end{figure}

We let $\theta_i$ for $i=L,R$ be the angles the $x_i$ axes  on the
left (L) and right (R) sides make with the normal to a straight GB
interface, Fig. 1.  GL treatments of the  Josephson current $I$
for a $d_{x^2-y^2}$-wave OP yielded for ideal (non-reconstructed)
and faceted GB's,\cite{SigristRice,TKYBCO}
\begin{eqnarray}
I^{i}_d&=&I^{i}_{cd}\cos(2\theta_L)\cos(2\theta_R)\sin(\delta\phi),\label{dGL}\\
I^f_d&=&I^{i}_{cd}\cos[2(\theta_L+\theta_R)]\sin(\delta\phi)/2,\label{randomd}
\end{eqnarray}
where $\delta\phi=\phi_L-\phi_R$ is the OP phase difference across
the GB.  These have the form $I=I_{c0}\sin(\delta\phi)$, where
$I_c=|I_{c0}|$ is the critical current. $I_{c0}>0$ and $I_{c0} <
0$ are the ``0-junction''
 and ``$\pi$-junction'' cases,  as $I=I_c\sin(\delta\phi+\pi)$ in the
latter.  There are three types of junctions we shall consider.
These are asymmetric, symmetric, and straight junctions.  We note
that other workers have used the terminology ``mirror'' and
``parallel'' for symmetric and straight junctions,
respectively.\cite{Shirai}
 For asymmetric junctions, $\theta_L=\theta,\theta_R=0$,
$I_d^{i,f}(\theta)/I_d^{i,f}(0)=\cos(2\theta)$.  For symmetric
junctions with $\theta_L=\theta_R=\theta$ and straight junctions
with $\theta_R=-\theta_L=\theta$,
$I_d^{i}(\theta)/I_d^{i}(0)=\cos^2(2\theta)$.   For faceted
symmetric and straight junctions,
$I_d^{f}(\theta)/I_d^f(0)=\cos(4\theta),1$, respectively. These
ideal and faceted GL results shown in Fig. 1 are in qualitative
agreement with our results for electron-doped cuprates with random
and specular tunneling, respectively, but differ qualitatively
from our microscopic results for hole-doped cuprates.

\section{Procedure}

To calculate  $I$ across an in-plane GB, we assume the Fourier
transform of the quasiparticle Green function matrix in the bulk
of the $i$th superconductor is\cite{ArnoldKlemm}
\begin{eqnarray}
\hat{\cal G}({\bm k}^i,\omega)&=&-\frac{i\omega+\Delta({\bm
k}^i)\tau_1+\xi({\bm k}^i)\tau_3}{\omega^2+|\Delta({\bm
k}^i)|^2+\xi^2({\bm k}^i)},\\
\xi_{0}({\bm
k}^i)&=&-J_{||}[\cos(k_x^ia)+\cos(k_y^ia)\nonumber\\
& &-\nu\cos(k_x^ia)\cos(k_y^ia)-\mu],\label{fermi}
\end{eqnarray}
where $\omega$ is a Matsubara frequency, ${\bm k}^i$ is the  wave
vector on the $i$th GB side, $\Delta({\bm k}^i)$ is the respective
two-dimensional (2D) gap function, $a$ is the tetragonal in-plane
lattice constant, $\mu$, $J_{||}$, and $\nu$  are parameters
defining the chemical potential, and in-plane bandwidth and
dispersion, respectively, and the $\tau_j$ are the Pauli matrices.
$\hat{\cal G}$ is the usual rank 2 matrix with elements $G, F,
-G^{\dag}$, and $F^{\dag}$ that describes both the single
quasiparticle and mean-field pair excitations.  We take $\xi({\bm
k}^i)=\xi_{0}({\bm k}^i)+J_{\perp}[1-\cos(k_z^is)]$, where $s$ is
the $c$-axis repeat distance. Here we are mainly interested in the
2D limit $J_{\perp}=0$.  For one tight-binding 2D FS appropriate
for Bi2212, FS2, we take $J_{||}=500$ meV, $\mu=0.6$, and
$\nu=1.3$, and set $\xi_{0}({\bm k}^i_F)=0$.\cite{ArnoldKlemm} A
slightly different FS, denoted ARPES, that most closely resembles
the FS of Bi2212 as observed in ARPES experiments, is obtained
with $J_{||}=306$ meV, $\mu=0.675$, and $\nu=0.90$.\cite{bks} A
nearly circular
 FS, FS3, is obtained with $\mu=1.0$ and
$\nu=0$.\cite{ArnoldKlemm}  Finally, we also studied the extreme
hole-doped near-neighbor tight-binding FS used by Shirai {\it et
al.}\cite{Shirai} Taking the maximum value of the gap to be 20
meV, this FS is governed by the parameters $J_{||}=201$ meV,
$\nu=0$, and $\mu=-1.94$. These four FS's are pictured in Fig. 2.

\begin{figure}
\includegraphics[width=0.45\textwidth]{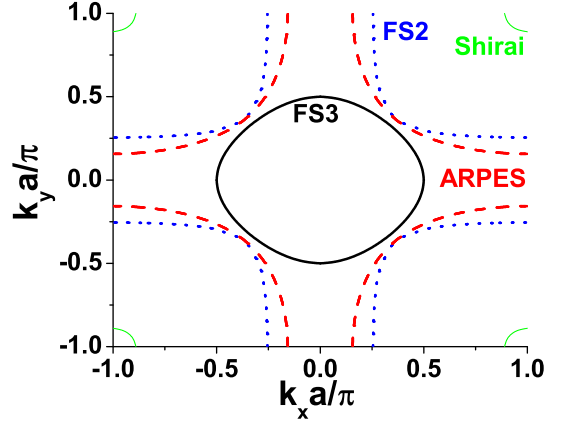}
\caption{(Color online) Plots of the four 2D Fermi surfaces
studied: The nearly isotropic FS3 (thick solid
black),\cite{ArnoldKlemm} the holed-doped FS2 (dotted
blue),\cite{ArnoldKlemm} the best fit to the ARPES FS of Bi2212
(dashed red,\cite{bks} and the extreme hole-doped FS studied by
Shirai (thin solid green).\cite{Shirai}} \label{fig2}
\end{figure}

Next, we construct the half-space Green function matrices on the
$i$th side of a
 straight (001) tilt GB.
We let $\hat{\bm n}$ and $\hat{\bm e}$ be unit vectors normal and
parallel to the GB satisfying $\hat{\bm e}\times\hat{\bm
n}=\hat{\bm z}$, as sketched in Fig. 1. Letting $k^i_{||}={\bm
k}_i\cdot\hat{\bf e}$ and $k^i_{\perp}={\bm k}_i\cdot\hat{\bm n}$,
we  set ${\bf k}^i_{||}=(k^i_{||},k_z^i)$,
$k_x^i=k^i_{\perp}\cos\theta_i-k^i_{||}\sin\theta_i$, and
$k_y^i=k^i_{\perp}\sin\theta_i+k^i_{||}\cos\theta_i$. The lack of
translational invariance  requires a discrete indexing of  the
lattice planes along $\hat{\bm n}$, as in our previous technique
for $c$-axis tunneling.\cite{ArnoldKlemm}. Thus, for integers $n,
m$,
\begin{eqnarray}
\hat{\cal G}_{m-n}({\bm
k}^i_{||},\omega)&=&\int_{-\pi/a^i_{\perp}}^{\pi/a^i_{\perp}}
\frac{dk^i_{\perp}}{2\pi}e^{ik^i_{\perp}a^i_{\perp}(m-n)}\hat{\cal
G}({\bm k}^i,\omega),
\end{eqnarray}
where ${\bm k}_{||}^i=(k_{||}^i,k_z^i)$ and the
$a^i_{\perp}/a={\rm min}(\cos\theta_i,\sin\theta_i)$ for
$90^{\circ}>\theta_i>0^{\circ}$, $a^i_{\perp}=a$ for
$\theta_i=0^{\circ},90^{\circ}$ are the respective  lattice plane
 separations along $\hat{\bm n}$. We choose the GB interface to be
between lattice planes 1 and 0 and set $n,m\ge1$, $n,m\le0$ for
$i=R,L$, respectively. Suppressing the  ${\bm k}^i_{||}$ and
$\omega$ dependencies, we construct each half-space $\hat{g}_{nm}$
from a combination of the $\hat{\cal G}_{mn}$ that obeys the free
surface boundary conditions (BC's) $\hat{g}_{m0}=\hat{g}_{0n}=0$
for $n,m\ge1$ and $\hat{g}_{m1}=\hat{g}_{1n}=0$ for  $n,m\le0$,
obtaining\cite{ArnoldKlemm,LiuKlemm}
\begin{eqnarray}
\hat{g}_{mn}&=&\hat{\cal G}_{|m-n|}-\hat{\cal
G}_{m+n},\qquad\>\>\>\>{\rm
for}\>m,n\ge1,\label{gmnp}\\
\hat{g}_{mn}&=&\hat{\cal G}_{|m-n|}-\hat{\cal
G}_{2-m-n},\qquad{\rm for}\>m,n\le0.\label{gmnm}
\end{eqnarray}

We then paste together the two half-space layered superconductors.
 The
supercurrent $I$ across the GB between the two half-spaces is
given for a general junction hopping matrix element $J_J({\bm
k}^L_{||},{\bm k}^U_{||})$ by
\begin{eqnarray}
I&=&\frac{ieT}{2}\sum_{{\bm k}^L_{||},{\bm
k}^U_{||},k_z,\omega}{\rm
Tr}\biggl[(\tau_0+\tau_3)\nonumber\\
&
&\times\biggl(J_J({\bm k}^L_{||},{\bm k}^U_{||})\hat{G}_{10}({\bm k}^U_{||},{\bm k}^L_{||},k_z,\omega)\nonumber\\
& &-J_J({\bm k}^U_{||},{\bm k}^L_{||})\hat{G}_{01}({\bm
k}^L_{||},{\bm k}^U_{||},k_z,\omega)\biggr)\biggr],\label{Iexact}
\end{eqnarray}
where $\hat{G}_{nm}({\bm k}^L_{||},{\bm k}^U_{||},k_z,\omega)$ is
the full Green's function matrix for the combined two half spaces
coupled via  tunneling across the (001) tilt GB.\cite{ArnoldKlemm}
An example of the resulting interface is pictured in Fig. 3.  In
principle, it is straightforward to evaluate $I$ to all orders in
$J_J$, as was done for the case of coherent tunneling across a
$c$-axis twist junction.\cite{ArnoldKlemm}  As in that case,
however, the most relevant case, especially for GB misalignment
angles $\theta_R+\theta_L$ sufficiently large, is for weak
tunneling across the GB.\cite{Mannhartreview}

\begin{figure}
\includegraphics[width=0.45\textwidth]{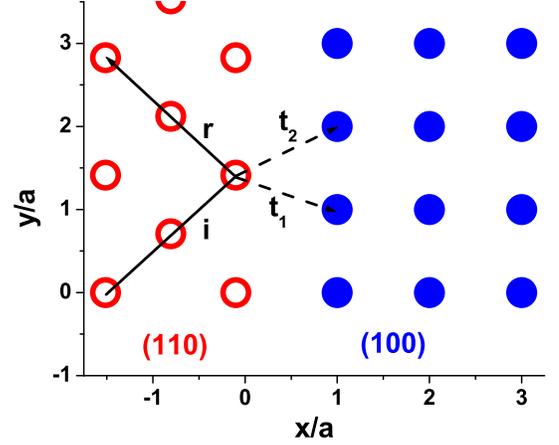}
\caption{(Color online) Sketch of a small part of a $(110)|(100)$
junction.  The bottom rows of the left $(110)$ lattice (open red
circles) and right $(100)$ (filled blue circles) lattice are
aligned.  An incident quasiparticle can hop along the path $i$,
and be specularly reflected along the path $r$, but tunnels
non-specularly into the $(100)$ lattice via the two primary
initial paths $t_1$ and $t_2$, as indicated.}\label{fig3}
\end{figure}

We set $\Delta({\bf k}^i)=\exp(i\phi_i){\rm Re}\Delta({\bf k}^i)$
and define $\overline{g}_{nm}^{pq}$ to be the $(pq)$th matrix
element of the rank 2 matrix $\hat{g}_{nm}$ with $\phi_i=0$.  To
leading order in the general tunneling matrix element $J_J({\bf
k}^L_{||},{\bf k}^R_{||})$, we obtain
\begin{eqnarray}
I&=&\frac{4eT}{A^2}\sum_{{\bf k}^R_{||},{\bf
k}^L_{||},\omega}|J_J({\bf k}^L_{||},{\bf k}^R_{||})|^2
\overline{g}_{11}^{12}({\bf k}^R_{||},\omega)\nonumber\\
& &\qquad\qquad\times\overline{g}_{00}^{21}({\bf
k}^L_{||},\omega)\sin(\delta\phi),\label{current}
\end{eqnarray}
where $A$ is the junction area, $e$ is the electron charge,
$\delta\phi=\phi_L-\phi_R$, $-\pi/s\le k^i_z\le\pi/s$,
$-\pi/a^i_{||}\le k^i_{||}\le\pi/a^i_{||}$, and the
$a^i_{||}=a^2/a^i_{\perp}$ are the lattice constants along
$\hat{\bm e}$.  We note that $\overline{g}_{11}^{12}$ and
$\overline{g}_{00}^{21}$ are the $F$ and $F^{\dag}$ pair functions
for real OP's on opposite sides of the junction, respectively.

We note that although Eq. (\ref{current}) is proportional to
$\sin\delta\phi$ since it is obtained in the limit of weak GB
tunneling, our exact expression for the supercurrent, Eq.
(\ref{Iexact}), also contains the full temperature dependence of
all of  the harmonics proportional to $\sin(n\delta\phi)$ for
$n\ge1$.

Experimentally for large symmetric  tilt angles
$\theta>\theta_c\approx5-7^{\circ}$, $I_c(\theta)\ll
I_c(0)$,\cite{Mannhartreview,Cai,Chisholm} justifying our weak
tunneling assumption, Eq. (\ref{current}), although the
approximation is less accurate for  $\theta<\theta_c$.  Others
working on (001) tilt grain boundary junctions of HTSC have also
made this assumption,\cite{Barash,YanHu} although they used a
circular FS and did not necessarily take account of the surface
boundary conditions.\cite{YanHu} Since the effects induced by the
GB to the real-space $d_{x^2-y^2}$-wave pairing interaction are
formidable, here we only treat the surface BC effects, and discuss
the effects of self-consistency elsewhere.\cite{ArnoldKlemmnext}
We note that a self-consistent calculation by Tanaka and Kashiwaya
for an OP of mixed symmetry suggested no qualitative changes upon
the imposition of self-consistency.\cite{TK1998} We include the
limiting cases of specular (or coherent) tunneling, where ${\bm
k}^R_{||}={\bm k}^L_{||}$, and random (or incoherent) tunneling,
where ${\bm k}^R_{||}$ and ${\bm k}^L_{||}$ are independent,
writing\cite{kars}
\begin{eqnarray}
|J_J({\bm k}^L_{||},{\bm k}^R_{||})|^2&= &|J_J^{\rm
sp}|^2\delta_{{\bm k}^R_{||},{\bm k}^L_{||}}A+|J_J^{\rm r}|^2.
\end{eqnarray}

We remark that the specular (coherent) and random (incoherent)
tunneling processes are microscopic processes that take place on
an atomic scale.  For specular tunneling to occur, the component
of the wave vector of the quasiparticle parallel to the junction
does not change at each atomic site upon tunneling.  With random
tunneling, the components of the quasiparticle wave vectors
parallel to the junction on the two sides of the junction are
completely independent variables. For example, let us consider a
tight-binding model of near-neighbor hopping   in a tetragonal
lattice, as in Eq. (4) with $\nu=0$. In this model, a
quasiparticle far from the junction hops from one site to an
adjacent site. In  a defect-free straight (or symmetric)
$(100)|(100)$ junction,  the lattices on both sides of the
junction are identical.  Thus,   a quasiparticle can propagate
from any site on one side of the junction to the nearest neighbor
site on the other side without changing its direction.  A
quasiparticle can also backscatter coherently at the interface,
preserving the component of the wave vector parallel to the
junction, as in other treatments.\cite{TanakaKashiwaya,Barash} For
instance, when the interface is a (110) surface, the surface is
theoretically atomically flat, allowing a substantial amount of
coherent reflection, as sketched in Fig. 3. Although with the
usual vapor phase epitaxy, such $(100)|(110)$ junctions are far
from perfect,\cite{Mannhartreview} when they were constructed
using the far superior liquid phase epitaxy, an excellent fit to
the Fraunhofer pattern was obtained by application of a magnetic
field parallel to the junction, without any evidence of a
zero-bias conductance peak at the interface,\cite{Eltsev} at least
in some cases. Such behavior is consistent with the incoherent
tunneling model of Ambegaokar and Baratoff (AB) for an $s$-wave
superconductor.  In earlier vapor deposition of the somewhat
different $(110)|(001)$ GB's, sometimes a rough approximation to
the Fraunhofer pattern was observed, with a peak at low fields,
and sometimes a very different pattern with a dip at zero field,
possibly consistent with a $d$-wave OP,\cite{YanHu} was
observed.\cite{Ishimaru}  In both of these experiments, high
resolution transmission electron microscopy (HRTEM) demonstrated
the quality of the junctions.\cite{Eltsev,Ishimaru}

However, at interfaces such as those containing a (130) surface,
Fig. 4, the amount of coherent reflection off the surface is
likely to depend strongly upon the incident quasiparticle wave
vector direction, as the surface is a periodic, but uneven zigzag.
To date, such junctions have never been prepared with the atomic
precision of the above two cases.\cite{Mannhartreview,Klemmreview}

\begin{figure}
\includegraphics[width=0.45\textwidth]{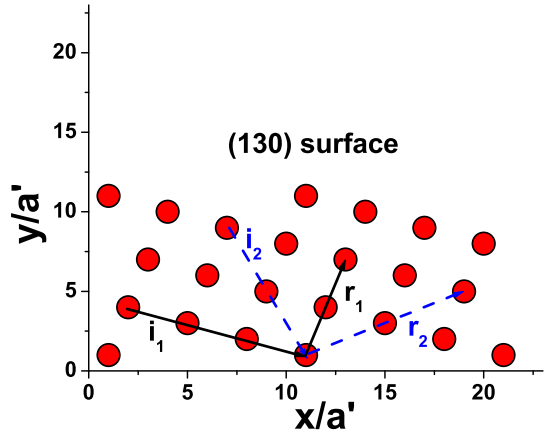}
\caption{(Color online) Sketch of the atomic sites (filled red
circles) of a tetragonal plane near a (130) surface (the $x$
axis), where $a'=\sqrt{10}a$.  The solid black and dashed blue
arrows indicate selected incident and reflected quasiparticle
paths involving only near-neighbor and next-nearest neighbor
hopping, respectively.}\label{fig4}
\end{figure}

 On the other hand, for an asymmetric $(100)|(110)$ junction, Fig. 3, the ratio of the parallel lattice
  spacings  on the two sides of the junction is $\sqrt{2}$, an irrational
  number.  Although specular reflection at the interface by a quasiparticle is entirely possible, in this case, even if the junction were atomically
  perfect,
  the proportion of atomic sites  at which
  coherent tunneling across the junction can  occur is vanishingly
  small.  That is, in tunneling to the nearest site across the
  junction, nearly all quasiparticles will change the component of their direction parallel to the
  junction.   One example of this is sketched in Fig. 3.  When one
  includes the dominant junction tunneling from all sites along the junction, the
   tunneling component of the quasiparticle wave vector parallel to
  the junction is effectively random.

  In general, one  expects that atomically perfect
  straight junctions can exhibit specular, or coherent tunneling.
  Defects would lead to a component of the tunneling that is
  random, or incoherent.
 Asymmetric junctions  and symmetric junctions that cannot also be described as straight
  junctions (such as the  $(100)|(100)$  junction) intrinsically lead to predominantly random, or
 incoherent tunneling.  Thus, atomically perfect $(pq0)|(100)$ asymmetric or $(pq0)|(pq0)$ symmetric
 junctions
 for $p+q\ne1$
 (with $\theta_L=\theta_R$) are expected to exhibit primarily
 random, or incoherent microscopic tunneling processes.  Facets,
 atomic disorder, stoichiometry variations, etc., at the junction
  contribute further to the randomness of the tunneling processes.\cite{Cai,Chisholm,Miller,Gurevich,Lang,Coppens,Moss,Grebille}

  This microscopic (atomic scale) coherence or incoherence of the tunneling process is completely
unrelated to the macroscopic coherence of the superconducting wave
function, which can occur over the entire area of the junction,
provided that it is prepared sufficiently uniformly. For example,
the AB microscopic model of Josephson tunneling,\cite{AB} which
averages independently over the parallel components of the
quasiparticle wave vector on each side of the junction, assumes
uniformly random, or incoherent, microscopic quasiparticle
tunneling only.  However, that model
 leads to macroscopic coherence of the superconducting wave
function on each side of the junction.  Note that although most
experiments are consistent with Bi2212 having strongly incoherent
intrinsic $c$-axis tunneling between the double layers, the
product of the critical current and the normal state resistance
$I_cR_n$ is about 20\% of that expected for a pure $s$-wave
superconductor with perfect interfaces, the AB
value.\cite{Klemmreview,AB} This occurs both for intrinsic and for
$c$-axis twist junctions several $\mu$m in
cross-section.\cite{Klemmreview,KleinerMoessle,Takano,Latyshev}

\section{Results}

\begin{figure}\includegraphics[width=0.45\textwidth]{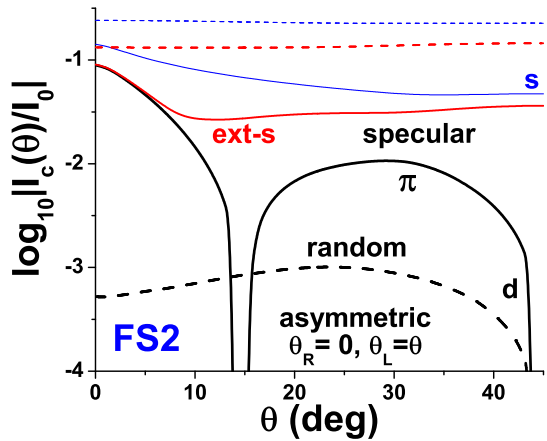}
\caption{(Color online) Plots of $\log_{10}|I_c(\theta)/I_0|$ for
asymmetric junctions with $\theta_R=0$, $\theta_L=\theta$, for
$s$- (thin blue), extended-$s$- (intermediate thickness red), and
$d_{x^2-y^2}$-wave (thick black) superconductors, for specular
(solid) and random (dashed) GB tunneling and the tight-binding
hole-doped FS2. The $d$-wave curves satisfy
$I_{c0}(90^{\circ}-\theta)=-I_{c0}(\theta)$, but the $s$- and
extended-$s$-wave curves satisfy
$I_{c0}(90^{\circ}-\theta)=I_{c0}(\theta)$.  See
text.}\label{fig5}
\end{figure}

\begin{figure}\includegraphics[width=0.45\textwidth]{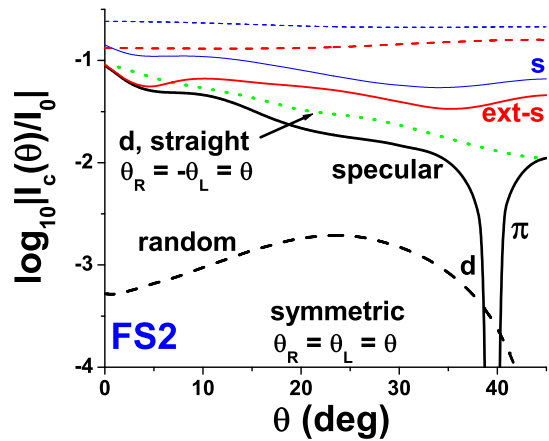}
\caption{(Color online) Plots of $\log_{10}|I_c(\theta)/I_0|$ for
symmetric junctions with $\theta_R=\theta_L=\theta$, for $s$-
(thin blue), extended-$s$- (intermediate thickness red), and
$d_{x^2-y^2}$-wave (thick black) OP's for specular (solid) and
random (dashed) GB tunneling with the tight-binding, hole-doped
FS2. Results for straight junctions with
$\theta_R=-\theta_L=\theta$ and a $d_{x^2-y^2}$-wave OP for
specular (dotted green) and random (also dashed black) GB
tunneling are also shown. All curves satisfy
$I_{c0}(90^{\circ}-\theta)=I_{c0}(\theta)$.}\label{fig6}
\end{figure}

In Fig. 5, we  present our  results obtained from Eq.
(\ref{current}) for the low-$T$    $I_c$ across asymmetric
junctions, $\theta_R=0$, $\theta_L=\theta$, including the
tight-binding  FS2 and the
 surface BC effects.  In all of our calculations, we
assume the $d_{x^2-y^2}$- and extended-$s$-wave OP's  have the
bulk real-space pairing forms
$\Delta_0(T)[\cos(k_x^ia)-\cos(k_y^ia)]$ and
$\Delta_0(T)|\cos(k_x^ia)-\cos(k_y^ia)|$, respectively.  The
ordinary-$s$-wave OP is  $\Delta_0(T)$, independent of ${\bm
k}^i$.\cite{ArnoldKlemm} We take $\Delta_0(0)=10$ meV, which
implies a maximum $d$-wave OP of 20 meV.  In Fig. 4, we plotted $
\log_{10}|I_c(\theta)/I_0|$, where $I_c=|I_{c0}|$ and
$I_0=4e|J_J^{sp}|^2/[a^3s(0.1J_{||})^2],
4e|J_J^r|^2/[a^4s^2(0.1J_{||})^2]$, respectively.\cite{kars} The
solid (dashed) curves are the results for specular (random)
tunneling, respectively. The specular $d$-wave GB junction behaves
as a $\pi$-junction only for $14.9^{\circ}\le\theta\le45^{\circ}$
and 75.1$^{\circ}\le\theta\le90^{\circ}$, qualitatively different
from the GL results pictured in Fig. 1.\cite{SigristRice} Note
that the $d$-wave $I_c$ for random GB tunneling is small but
non-vanishing due to  the surface BC's. For this and all other
cases of random tunneling across GB's between $d_{x^2-y^2}$-wave
superconductors, neglect of the surface BC's would lead
necessarily to $I_c=0$.  In the 2D limit with random GB tunneling,
however, the asymmetric GB results are qualitatively in agreement
with the GL results, except for the overall magnitude of $I_c$,
which is greatly reduced from that obtained for specular GB
tunneling.

We  also investigated the role of $J_{\perp}$ for the $d$-wave
case.  For random GB tunneling, as $J_{\perp}$ increases to 40
meV, the broad peak in $I_{c0}(\theta)$ becomes much flatter, and
for much larger $J_{\perp}$, $I_{c0}(\theta)$ non-monotonically
approaches the GL result $\propto\cos(2\theta)$.  For specular
tunneling, $J_{\perp}=10$ meV yields $d$-wave results nearly
identical to those pictured for $J_{\perp}=0$.  Increasing
$J_{\perp}$ beyond $J_{\perp0}\approx$40 meV eliminates the
regions within $0\le\theta\le45^{\circ}$ of $\pi$-junctions, so
that the overall region of $\pi$-junctions is for
$45^{\circ}\le\theta\le90^{\circ}$, as for the GL case.  Hence,
for $J_{\perp}\le J_{\perp0}$, the specular, tight-binding
asymmetric GB $d$-wave results with FS2 are qualitatively
different than the GL ones.

In Fig. 6, we present the analogous results for symmetric
junctions with FS2.  We also plotted our results for straight GB
junctions with a $d_{x^2-y^2}$-wave OP.  For random GB tunneling,
the straight GB junction results are identical to the dashed black
curve for random GB tunneling with symmetric junctions.  For
specular GB tunneling, the straight junction results are shown as
the dotted green curve.
 Note that for specular tunneling, straight and symmetric junctions
 have opposite $I_c$ values at $\theta=45^{\circ}$.  For symmetric GB junctions,
 the specular $d$-wave junction behaves as a
$\pi$-junction for $39.4^{\circ}\le\theta\le50.6^{\circ}$, but
behaves as a 0-junction otherwise.   For random tunneling, the
symmetric $d$-wave GB junction behaves similarly to the ideal GL
model, Eq. (\ref{dGL}), as both have $I_c(45^{\circ})=0$ but never
behave as $\pi$-junctions. However, the $d$-wave GL facet model,
Eq. (2), has $I_d^f(45^{\circ})=-I_d^f(0^{\circ})$, which is
qualitatively different.\cite{TKYBCO} We note from Figs. 5, 6 that
for small angle ($0\le\theta\le5^{\circ}$) specular
 GB junctions, the extended-$s$ and
$d_{x^2-y^2}$-wave OP's lead to indistinguishable $I_c(\theta)$.
However, for random GB tunneling, the $d$-wave $I_c(\theta)$ is
much smaller than the extended-$s$-wave $I_c(\theta)$, and would
vanish in the absence of the surface BC's.  Except for the broad
peak and the greatly reduced magnitude of $I_c$, the symmetric GB
junction across $d_{x^2-y^2}$-wave superconductors with FS2 and
random GB tunneling is qualitatively in agreement with the GL
result of Fig. 1.\cite{SigristRice}  For specular GB tunneling,
however, our microscopic $d_{x^2-y^2}$ results differ
qualitatively from the GL forms.\cite{SigristRice}

\begin{figure}
\includegraphics[width=0.45\textwidth]{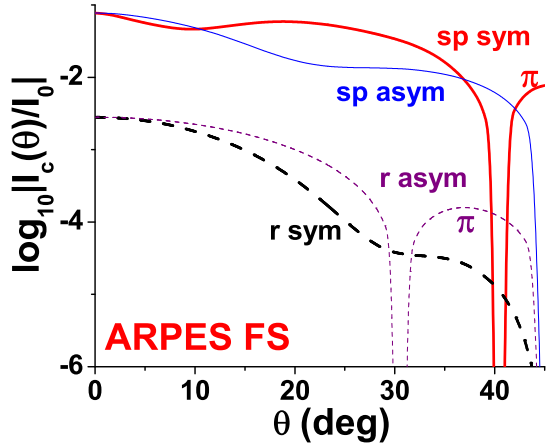} \caption{(Color online) Plots
of $\log_{10}|I_c(\theta)/I_0|$ for symmetric junctions with
specular (thick solid red) and random (thick dashed black)
tunneling, and asymmetric junctions with specular (thin solid
blue) and random (thin dashed purple) tunneling, for a
$d_{x^2-y^2}$-wave OP with the ARPES FS.\cite{bks}  The straight
and symmetric GB junctions satisfy
$I_c(90^{\circ}-\theta)=I_c(\theta)$, but the asymmetric GB
junctions satisfy
$I_c(90^{\circ}-\theta)=-I_c(\theta)$.}\label{fig7}
\end{figure}

We remark that the ordinary-$s$- and extended-$s$-wave OP's lead
to $I_c(0)$  values that are comparable to each other for both GB
tunneling mechanisms.  However, for the $d_{x^2-y^2}$-wave OP,
random GB tunneling with FS2 leads to a suppression of $I_c(0)$ by
a factor of 461 from that of the ordinary-$s$-wave OP. This
suggests that $I_c(0)$ for the $d_{x^2-y^2}$-wave OP with weak,
random GB tunneling is unlikely to be consistent with the large
values observed experimentally across low-angle (001) tilt
junctions.\cite{Mannhartreview}  The broad peak in
$I_{c0}(\theta)$ for the $d$-wave OP with random GB tunneling for
both asymmetric and symmetric junctions is also inconsistent with
most experimental data.  On the other hand, the $d$-wave OP with
specular GB tunneling across asymmetric junctions does show a
rapid decrease in $I_{c0}(\theta)/I_{c0}(0)$ with increasing
$\theta$, in agreement with low-angle
experiments,\cite{Mannhartreview} but vanishes at 14.9$^{\circ}$
and 39.4$^{\circ}$ for FS2, respectively, unlike the
experiments.\cite{Mannhartreview}

In Figs. 7-9, we compare our $\log_{10}|I_c(\theta)/I_0|$ results
for $d_{x^2-y^2}$-wave superconductors across symmetric and
asymmetric GB junctions with three different FS's. For each FS,
$I_c$ results for straight and symmetric GB junctions with random
GB tunneling are identical, and for specular GB tunneling, the
$I_c(45^{\circ})$'s are opposite in sign and equal in magnitude.
In Figs. 7-9, the results for symmetric GB junctions with specular
(random) GB tunneling are shown as the thick solid red (dashed
black) curves, respectively, and results for asymmetric GB
junctions with specular (random) GB tunneling are shown as the
thin solid blue (dashed purple) curves, respectively. Results for
straight and symmetric GB junctions with random tunneling are
identical, and the results for straight GB junctions with specular
GB tunneling are shown as the dotted green curves.

In Fig. 7 we plotted $\log_{10}|I_c(\theta)/I_0|$ versus $\theta$
for symmetric and asymmetric GB junctions for a $d_{x^2-y^2}$-wave
superconductor with the ARPES FS. Note that for symmetric
junctions, a $\pi$-junction occurs with specular GB tunneling for
$40.3^{\circ}\le\theta\le49.7^{\circ}$,  very similar to the
region obtained for FS2 pictured in Fig. 5. In addition, for
random GB tunneling across symmetric junctions, no $\pi$-junctions
are obtained.  However, for asymmetric junctions, the results
differ substantially from those obtained for a $d_{x^2-y^2}$-wave
superconductor with FS2, pictured in Fig. 4. In that (FS2) case,
the asymmetric GB junction was a $\pi$-junction for specular GB
tunneling with $14.9^{\circ}<\theta<45^{\circ}$ and
$75.1^{\circ}<\theta\le90^{\circ}$, but was  a 0-junction
otherwise.  For random GB tunneling, it was always a 0-junction
for $0\le\theta\le45^{\circ}$ and a $\pi$-junction for
$45^{\circ}\le\theta\le90^{\circ}$. For this slightly different
(see Fig. 2) ARPES FS, however, nearly the opposite behavior was
found. Asymmetric GB junctions with the ARPES FS with specular GB
tunneling are 0-junctions
 for $0\le\theta<45^{\circ}$ and $\pi$-junctions for
 $45^{\circ}<\theta\le90^{\circ}$, but for random tunneling, a $\pi$-junction appears for
$30.6^{\circ}<\theta<45^{\circ}$ (and also for
$59.4^{\circ}<\theta\le90^{\circ}$), and a 0-junction otherwise.
Hence, these slightly different hole-doped FS's lead to
qualitatively different $I_c$ behavior for $d_{x^2-y^2}$-wave
superconductors. For $s$- and extended-$s$-wave superconductors,
on the other hand, such strong FS dependence of $I_c(\theta)$ does
not occur. Although for clarity in presenting the $d$-wave
results, we have not shown our $s$- and extended-$s$-wave data for
the ARPES FS, our calculations confirm the close similarity of
those results to the analogous ones pictured in Figs. 5 and 6,
confirming the  reliability of our results.

With the ARPES FS, asymmetric GB junctions with specular tunneling
and for symmetric GB junctions with random tunneling are
qualitatively consistent with the ideal GL
model,\cite{SigristRice} and symmetric GB junctions with specular
tunneling are qualitatively consistent with the GL facet
model,\cite{TKYBCO} except that the position of $I_c(\theta)=0$
has shifted from $22.5^{\circ}$ to $40.3^{\circ}$. However,
asymmetric GB junctions with random tunneling and the ARPES FS are
completely unlike either  GL model.

Next, we calculated $I_{c0}(\theta)$ at low $T$ across GB's with
the nearly circular FS3, appropriate for the electron-doped
cuprates such as NCCO, and presented our results in Fig. 8.  For
asymmetric GB's, our $d$-wave results agree with the GL model,
$I_{c0}(\theta)/I_{c0}(0)\approx\cos(2\theta)$, for both specular
and random GB tunneling, but for the latter, $I_{c0}(0)$ is
greatly reduced in magnitude for a $d$-wave OP. Our results for
symmetric or straight GB's with random GB tunneling coincide with
each other, and $I_{c0}(\theta)/I_{c0}(0)\approx\cos^2(2\theta)$,
as in the GL model for ideal symmetric GB junctions (Fig. 1).  For
specular GB tunneling across symmetric GB junctions, the results
are qualitatively similar to the GL facet model, Eq. (2) except
that the  $\theta$ value for the crossover from $0$- to
$\pi$-junction behavior has shifted from $22.5^{\circ}$ to
29.8$^{\circ}$. With that modification, all of our $d$-wave
results for the electron-doped FS3 are qualitatively in agreement
with the GL facet model for specular GB tunneling,\cite{TKYBCO}
and with the GL ideal model for random GB tunneling between
$d_{x^2-y^2}$-wave superconductors.\cite{SigristRice}  This
conclusion is somewhat unexpected, as one na{\"\i}vely might
 have expected the specular and random results with
FS3 to agree with the ideal and facet GL models, respectively. Our
low-$T$ symmetric junction results are also qualitatively in
agreement with those obtained at low $T$ for weak tunneling with a
perfectly circular FS and a FS-restricted $d$-wave OP with
different GB roughness,\cite{Barash} and our straight GB junction
results with random GB tunneling are qualitatively consistent with
previous straight, dirty-$N$, SNS junction results.\cite{Asano}

\begin{figure}
\includegraphics[width=0.45\textwidth]{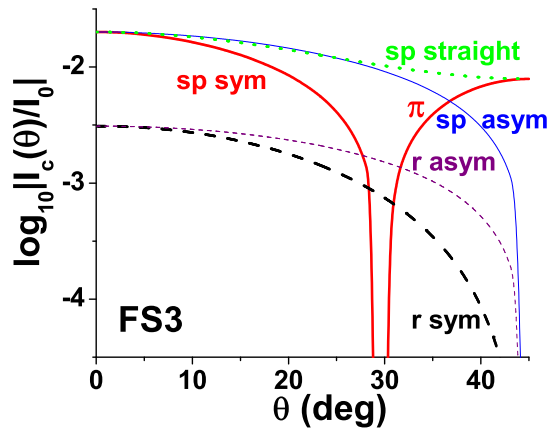} \caption{(Color online) Plots
of $\log_{10}|I_c(\theta)/I_0|$ for symmetric junctions with
specular (thick solid red) and random (thick dashed black)
tunneling, and asymmetric junctions with specular (thin solid
blue) and random (thin dashed purple) tunneling, for a
$d_{x^2-y^2}$-wave OP with the nearly isotropic,
``electron-doped'' FS3.  Also shown is the specular $d$-wave curve
(dotted green) for straight junctions. The straight and symmetric
junctions satisfy $I_c(90^{\circ}-\theta)=I_c(\theta)$, but the
asymmetric junctions satisfy
$I_c(90^{\circ}-\theta)=-I_c(\theta)$.}\label{fig8}
\end{figure}

In order to compare the results obtained using our technique of
imposing the surface BC's upon the Green function matrix to that
obtained using different techniques by others, we performed
explicit calculations for these purposes. According to the
formulas in their paper, Shirai {\it et al.} used the FS pictured
by the small green pockets near the corners of the first BZ
pictured in Fig. 2.\cite{Shirai} Those authors constructed
half-spaces exhibiting (100), (130), (120), and (110) surfaces,
and pasted them together with specular tunneling in either the
straight or symmetric configurations (which they denoted as
parallel and mirror, respectively).  Their discrete quantization
axis was always along the $x$ direction parallel to the CuO bonds,
and hence only normal to the surface for the (100) case. The
number of rows in the $y$ direction varied from 1 for the $(100)$
and $(110)$ surfaces to 3 for the $(130)$ surface.  Unlike our
quantization axis that was defined to be parallel and normal to
the surface as pictured in Figs. 3 and 4, upon Fourier
transformation in the $y$ direction, their non-normal quantization
lattice basis set could lead to unintended hopping across the GB
with the strong, intrinsic tight-binding near-neighbor hopping
matrix element.

In addition, Shirai {\it et al.} included tunneling from a few
rows other than the surface row of idealized atomic sites, by
assuming that the specular GB tunneling matrix elements decreased
inversely instead of exponentially  with the tunneling distance,
taking the maximum hopping strength to be 0.05$t$, or 6.2\% of the
effective near-neighbor hopping strength, $0.803t$.\cite{Shirai}
Nevertheless, they calculated the single quasiparticle densities
of states and $I_c$ across the straight and symmetric GB junctions
with the above form of specular tunneling, for $\theta=0^{\circ},
18.4^{\circ}, 26.5^{\circ}$, and $45^{\circ}$, respectively. Some
of their low-$T$ limits of $I_c$  can be compared with ours using
their unusual FS.\cite{Shirai}  Although they did not mention it
explicitly, Shirai {\it et al.} found  the straight
$\theta=0^{\circ}$ and $26.5^{\circ}$,  junctions to be
0-junctions. However,  they found strong zero-bias conductance
peaks at all junction sites for the straight $45^{\circ}$
junction, and at two of the three distinct junction sites for
their $18.4^{\circ}$ junction, suggesting that both of these
junctions behave as $\pi$-junctions.   For the symmetric
junctions, they found the low-$T$ limits of the symmetric
18.4$^{\circ}$ and 45$^{\circ}$ junctions to be $\pi$-junctions,
but the low-$T$ limits of the $0^{\circ}$ and $26.5^{\circ}$
symmetric GB junctions to be 0-junctions.\cite{Shirai}  Of these,
$I_c(T)$ for the 18.4$^{\circ}$ symmetric junction changed sign at
a finite $T$ value.

Our low-$T$ results for $\log_{10}|I_c(\theta)/I_0|$ versus
$\theta$ for symmetric, asymmetric, and straight junctions with
the Shirai FS are pictured in Fig. 9.  In all cases, even for
$\theta=0$, the extremely small overlap of the Fermi surfaces for
specular GB tunneling leads to an extremely small $I_c$. Our
results for specular GB tunneling with symmetric junctions are
shown as the  thick solid red curves with three alternating
regions of $\pi$- and 0-junctions for $0\le\theta\le45^{\circ}$.
In the domain $0^{\circ}\le\theta\le45^{\circ}$, $\pi$-junctions
are found in the three regions $1.57^{\circ}<\theta<7.19^{\circ}$,
$9.85^{\circ}<\theta<20.89^{\circ}$, and
$31.52^{\circ}<\theta<45^{\circ}$.  The behavior of the domain
$45^{\circ}\le\theta\le90^{\circ}$ is obtained from
$I_c(90^{\circ}-\theta)=I_c(\theta)$.  Random tunneling across
either straight or symmetric GB junctions leads to just a
0-junction for $0\le\theta\le90^{\circ}$, although there  are
strong dips in $I_c$ at $\theta\approx13.6^{\circ}$ and
$76.4^{\circ}$) (not pictured) and $I_c(45^{\circ})=0$, as
indicated by the thick dashed black curve. For asymmetric GB
junctions with specular GB tunneling, the thin solid blue curve
shows a $\pi$-junction for $17.04^{\circ}<\theta<45^{\circ}$ (and
from $72.96^{\circ}<\theta\le90^{\circ}$), and the thin dashed
purple curve for random GB tunneling indicates a $\pi$-junction
for $13.34^{\circ}<\theta<45^{\circ}$ (and from
$76.66<\theta\le90^{\circ}$).  In addition, the dotted green curve
represents straight junctions with specular tunneling, which are
always 0-junctions, as in the ideal and faceted  GL models
pictured in Fig. 1, and in all other FS's we studied (e. g., Figs.
6 and 8).

\begin{figure}
\includegraphics[width=0.45\textwidth]{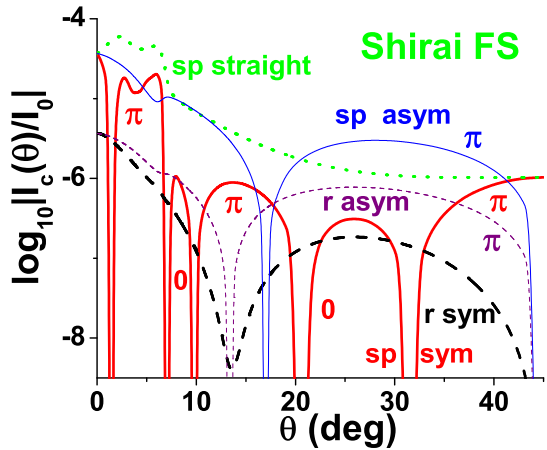} \caption{(Color online) Plots
of $\log_{10}|I_c(\theta)/I_0|$ for symmetric junctions with
specular (thick solid red) and random (thick dashed black)
tunneling, and asymmetric junctions with specular (thin solid
blue) and random (thin dashed purple) tunneling, for a
$d_{x^2-y^2}$-wave OP with the extreme hole-doped FS of Shirai
{\it et al.}\cite{Shirai} Also shown is the specular $d$-wave
curve (dotted green) for straight junctions. The straight and
symmetric junctions satisfy $I_c(90^{\circ}-\theta)=I_c(\theta)$,
but the asymmetric junctions satisfy
$I_c(90^{\circ}-\theta)=-I_c(\theta)$. The thick dashed black
curve represents  0-junctions only.}\label{fig9}
\end{figure}

Some of our results and those of Shirai can be directly compared.
Both procedures lead to $T=0$ $\pi$-junctions for symmetric GB
junctions with specular GB tunneling with $\theta=18.4^{\circ},
45^{\circ}$, but 0-junctions at $\theta=0^{\circ}, 26.5^{\circ}$.
As for straight junctions, our results for all four FS's we
studied always lead to 0-junctions, as for both the ideal and
faceted  GL models. For straight junctions at the two angles
$0^{\circ}$ and $26.5^{\circ}$,  Shirai {\it et al.} also obtained
0-junctions for all $T\le T_c$. However, as mentioned above,
Shirai {\it et al.} appear to have obtained  $\pi$-junctions for
the straight junctions with $\theta=18.4^{\circ}$ and
45$^{\circ}$.    In particular, we find their apparent
$\pi$-junctions for the two straight junctions,
$\theta=18.4^{\circ}$, $45^{\circ}$ to be difficult to understand.
Suppose Shirai {\it et al.}  were to increase their dominant
junction coupling strength from 0.05$t$ to their bulk hopping
strength, $0.803t$.  Then, the junction would be invisible to the
quasiparticles, and one would certainly not expect any surface
effects, such as in the straight 0$^{\circ}$ junction they
studied.\cite{Shirai} Although Shirai {\it et al.} attributed
their peculiar results to the oddness of the number of site rows
in their minicell, we suspect that their results may be artifacts
of their peculiar minicell basis, which was not defined to be
parallel and perpendicular to the GB, as in Fig. 4.\cite{Shirai}

Finally, in Fig. 10 we compare our specular
$I_{c0}(\theta)/I_{c0}(0)$ $d$-wave results with the surface BC's
with those obtained from the bulk Green functions. We compare
those results obtained with the tight-binding  FS2 for asymmetric
(Fig. 5) and symmetric (Fig. 6) junctions, and also our results
for symmetric junctions obtained with the nearly circular FS3
(Fig. 8).  For both symmetric GB cases, there are no regions in
$0^{\circ}\le\theta\le45^{\circ}$ of $\pi$-junctions for the bulk
calculations.   Including the surface effects, these symmetric GB
cases have $\pi$-junctions for $39.4^{\circ}<\theta<50.6^{\circ}$
 and
$29.2^{\circ}<\theta<55.8^{\circ}$ for FS2 and FS3, respectively.
For asymmetric junctions with  FS2, the bulk calculation leads to
 weak $\pi$-junctions for
$23.7^{\circ}<\theta<28.3^{\circ}$ with a negative slope at
$45^{\circ}$, yielding additional  $\pi$-junctions for
$45^{\circ}<\theta< 61.7^{\circ}$ and
$66.3^{\circ}<\theta\le90^{\circ}$. The surface BC's greatly
enhance the magnitude and range of such $\pi$-junctions to
$14.9^{\circ}<\theta<45^{\circ}$ and
$75.1^{\circ}<\theta\le90^{\circ}$, and change the sign of the
slope at 45$^{\circ}$. Although not pictured in Fig. 10, the
surface BC's play an essential role in the case of random
tunneling across GB's between $d$-wave superconductors.  In the
absence of the surface BC's, $I_c=0$ in such cases.  Thus, we
conclude that the surface BC effects cannot be ignored, especially
with regard to $d_{x^2-y^2}$-wave superconductors.

\begin{figure}
\includegraphics[width=0.45\textwidth]{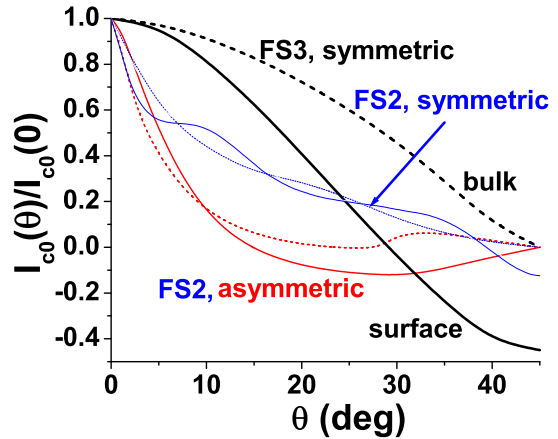}
\caption{(Color online) Plots of $I_{c0}(\theta)/I_{c0}(0)$ for
the $d_{x^2-y^2}$-wave OP with specular tunneling, with (solid)
and without (dashed) the surface boundary conditions, for
symmetric GB junctions with  the  nearly circular FS3 (thick
black) and the tight-binding FS2 (thin red) and , and for
asymmetric GB junctions with FS2 (intermediate thickness blue).
The symmetric curves satisfy $I_c(90^{\circ}-\theta)=I_c(\theta)$,
and the antisymmetric curves satisfy
$I_c(90^{\circ}-\theta)=-I_c(\theta)$.}\label{fig10}
\end{figure}

\section{Discussion}

Tricrystal experiments  were performed for a variety of cuprate
superconductors.\cite{TKYBCO,TKBi2212,TKhole,TKreview} Tricrystal
(a) consists of one asymmetric 0$^{\circ}/60^{\circ}$ junction and
two 0$^{\circ}/30^{\circ}$ junctions.  Tricrystal (b) consists of
two 0$^{\circ}/71.6^{\circ}$ junctions and one symmetric
18.4$^{\circ}/18.4^{\circ}$ junction. Tricrystal (c) consists of
two asymmetric 0$^{\circ}/73^{\circ}$ junctions and an asymmetric
2$^{\circ}/32^{\circ}$ junction.\cite{TKreview,TKNCCO}  It was
claimed that an odd number of $\pi$-junctions was observed only
for tricrystal (a).\cite{TKreview,TKNCCO}  In the case of Bi2212,
only results for tricrystal (a) were shown.\cite{TKBi2212,TKhole}

First of all, our results for the electron-doped FS3 with either
specular or random GB tunneling are in complete agreement with a
predominant $d_{x^2-y^2}$-wave OP as determined from the
tricrystal experiments on NCCO.\cite{TKNCCO} That is, one would
only expect an odd number of $\pi$-junctions only for tricrystal
(a), for both random and specular GB tunneling.

However, for hole-doped cuprates, the agreement with the
tricrystal experiments is  less robust. Assuming only a
$d_{x^2-y^2}$-wave OP,\cite{TKYBCO} our FS2 results pictured in
Figs. 3, 4 show that tricrystal (a) would have two $\pi$-junctions
if the tunneling were specular, and one $\pi$-junction if it were
random. Tricrystal (b) should have no $\pi$-junctions,
 as claimed. Tricrystal
(c), if the 2$^{\circ}/32^{\circ}$ junction can be approximated by
an asymmetric 0$^{\circ}/34^{\circ}$ junction, would have one
$\pi$-junction for specular tunneling, and two $\pi$-junctions for
random tunneling. Hence, our FS2 results are only consistent with
the observation claims if the tunneling is non-specular.

On the other hand, the ARPES FS, while only slightly different
from FS2, leads to very different conclusions with regard to the
agreement with the tricrystal experiments.  For specular GB
tunneling, the ARPES FS is in complete agreement with the
observations of the three tricrystal experiments.  However, for
random GB tunneling, tricrystal (a) is expected to have a very
small overall $I_c$, as $I_c\approx0$ for all three asymmetric GB
junctions in that tricrystal.  Although tricrystal (b) would be
expected to exhibit an even number of $\pi$-junctions, as claimed
to be observed, tricrystal (c) would be expected to exhibit one
$\pi$-junction for random GB tunneling with the ARPES FS.  Hence,
the ARPES FS can only be fully consistent with the tricrystal
experiments on hole-doped cuprates for non-random GB tunneling.

One interpretation is that if the quasiparticle tunneling across
the relevant grain boundaries is random and the OP were pure
$d$-wave, then the FS would be expected to be more like FS2. In
the unlikely event that the grain boundary tunneling were
coherent, then a pure $d$-wave OP would have a FS more like the
ARPES FS.  Such a scenario might be used to argue that the FS of
Bi2212, usually thought to have the ARPES FS form, is actually
closer to FS2. However, we caution that FS2 and the ARPES FS are
one-band approximations to the FS of Bi2212, which contains two
CuO$_2$ layers thought by many to be responsible for the
superconductivity, and that no tricrystal (c) experiments were
performed on Bi2212.  In addition, this scenario ignores the
$c$-axis twist experiments, which indicate that essentially all of
the $c$-axis tunneling is in the $s$-wave
channel.\cite{Klemmreview} If those three experiments are correct,
then Bi2212 contains an OP with an $s$-wave component that is at
least 20\% of the total. Although not appreciated by Tsuei {\it et
al.},\cite{TKhole} in the presence of nanoscale disorder that has
recently been observed in STM experiments,\cite{Lang} a mixed
$d+s$ OP is indeed allowed.\cite{Klemmreview}  Thus, it is
possible that Bi2212 could have a dominant $d$-wave OP and still
have all of the intrinsic $c$-axis tunneling be in the $s$-wave OP
channel.  If this scenario proves to be true, then further
calculations with a mixed $s$- and $d$-wave OP would be
warranted.\cite{Klemm}

With regard to the most-studied material, YBCO, in addition to the
two CuO$_2$ layers, there is also a conducting CuO chain layer,
which is responsible for the orthorhombicity and the strong
twinning that occurs in thin films.  The strong twinning makes
(100) and (010) films indistinguishable, which can complicate the
analysis.  Second, unless strongly underdoped, YBCO appears to be
more three-dimensional than are most other cuprates, so that
$J_{\perp}\ne0$ might be large enough to cause an important
modification to our results. Third, it is clear from many other
experiments that YBCO also has a substantial $s$-wave
component,\cite{Dynes,Clarke} which has often been ignored by many
workers.\cite{TKreview,TKhole}  In each twin domain, a purported
dominant $d$-wave OP would have either a $d+s$ or a $d-s$ OP mix,
and if the twin areas were equal, then one might expect the
$s$-wave OP component to make a negligible contribution to the
measured critical currents. However, in $c$-axis YBCO/Pb Josephson
junctions made with highly twinned YBCO thin films, quantitative
agreement to the Fraunhofer pattern was obtained, although
$I_cR_n$ was considerably less than the AB value.\cite{Katz}

Thus, the different results of FS2 and the ARPES FS indicate  that
the tricrystal experiments are  less robust determinations of the
symmetry of the order parameter in hole-doped cuprates than has
been generally thought. Small variations in the FS topology and
the non-specularity of the GB tunneling processes play a crucial
role that cannot be anticipated {\it a priori}.  To the extent
that defects play a role in the tunneling processes, it might be
possible to understand the experimental observations with a
substantial $s$-wave OP component. Especially since the (001) tilt
GB junctions are known to be increasingly underdoped with
increasing GB
misalignment,\cite{Mannhartreview,Cai,Chisholm,Gurevich} it is
possible that $\pi$-junctions arising from the local magnetism of
the Cu$^{+}$ spins formed from oxygen loss at the GB junctions
could be important.

All of our calculations using the Shirai FS are in complete
disagreement with both the tricrystal and tetracrystal
experiments, regardless of the degree of randomness in the GB
tunneling matrix elements. In addition, the Shirai {\it et al.}
results for specular symmetric 18.4$^{\circ}/18.4^{\circ}$ GB's
are in direct conflict with all experiments using tricrystal
(b).\cite{Shirai,TKreview}
 Taking all of our results for the various FS's into account, it
appears that there is a topological crossover relevant for the
agreement of the critical currents with the GL
results,\cite{SigristRice} with full agreement for the
electron-doped FS's centered about the $\Gamma$ point in the first
BZ, and complete disagreement with the extreme Shirai FS.  The
cross-over appears to occur for FS's intermediate to the ARPES FS
and FS2.  Hence, for the hole-doped cuprates, not only are the
details of the tunneling important, but the topological structure
of the hole-doped FS is also  very relevant, and minute details
can change the interpretation.

Tetracrystals consisting of highly twinned thin films of the
hole-doped YBCO and electron-doped LCCO containing a symmetric
$0^{\circ}/0^{\circ}$ junction and a 45$^{\circ}/\pm45^{\circ}$
junction, plus two symmetric $22.5^{\circ}/22.5^{\circ}$ junctions
were made.\cite{MannhartSQUID,Chesca} For comparison, they also
made bicrystals with two symmetric 22.5$^{\circ}/22.5^{\circ}$
junctions.  We  also calculated the straight GB junction case
$\theta_R=-\theta_L=\theta$ for a $d_{x^2-y^2}$-wave OP, and our
results for FS2 and FS3 are shown by the green dotted lines in
Figs. 6 and 8, and at 45$^{\circ}$, are opposite in sign to the
respective symmetric junction cases with specular GB tunneling. If
the tunneling were non-specular, as required for the $d$-wave
interpretation with FS2 of the tricrystal experiments, then the
45$^{\circ}/\pm45^{\circ}$ junction would have the smallest (or
vanishing, for random tunneling) $I_c$ value for a $d$-wave OP. If
instead the tunneling were specular, as required for the ARPES FS
$d$-wave interpretation of the tricrystal experiments on
hole-doped cuprates, then that junction would allow a small but
significant $I_c$ value, but would behave either as a $\pi$- or
0-junction, respectively, for a $d$-wave OP, depending upon the
particular distribution of twin domains present at the junction.
Thus, this tetracrystal would have an indeterminate number of
$\pi$-junctions for specular tunneling and a $d$-wave OP, but no
supercurrent for random tunneling.   Similar arguments apply to
the LCCO case.  For a highly twinned thin film junction, both
specular and random tunneling across the
 45$^{\circ}/\pm45^{\circ}$ GB junction would lead to $I_c\approx0$ for
 a $d_{x^2-y^2}$-wave OP, contrary
to the claims in the experiment.\cite{MannhartSQUID} The other two
junctions, symmetric  22.5$^{\circ}/22.5^{\circ}$ junctions, would
not be $\pi$-junctions, unless the tunneling were specular and the
FS were different from any that we studied. Hence, the
interpretation put forward by those authors, which assumed the
45$^{\circ}/\pm45^{\circ}$ junctions were straight,  should be
reexamined,\cite{MannhartSQUID,Chesca} as the $\pi$-junction could
arise extrinsically from defects in this very imperfect
45$^{\circ}/\pm45^{\circ}$
junction.\cite{Mannhartreview,Cai,Chisholm,Gurevich} We urge that
the experiment be redesigned in the form of a tricrystal, omitting
the 45$^{\circ}/\pm45^{\circ}$ junction.

Recently, there have been  zigzag in-plane YBCO/Nb
experiments.\cite{Smilde,HilgenkampNature}  In these experiments,
a YBCO film is angularly cut from above with a ramping angle
$\varphi\approx 15-20^{\circ}$ into a regular zigzag pattern of
nominally exposed (100) and (010) edges, each edge of equal length
$\ell\approx10-40\mu$m. After deposition of a thick layer of Au, a
thicker layer of Nb was applied.  This experiment is a variation
of the earlier $\theta=90^{\circ}$ YBCO/Pb superconducting quantum
interference device (SQUID) experiment of Mathai {\it et
al.},\cite{Mathai} which contained YBCO/Pb junctions with a thin
Ag interstitial layer along the ramped nominally (100) and (010)
edges of YBCO, respectively.  In those experiments, one of the
junctions functioned as a 0-junction, and the other functioned as
a $\pi$-junction.\cite{Mathai} The beauty of the recent experiment
is that the zigzag array of alternating 0-/$\pi$-junctions was
observed by a SQUID microscope to exhibit an ordered array of
fractional vortices,
 the magnitude of which was
within the experimental error of $\Phi_0/2$, where $\Phi_0$ is the
flux quantum.  Although no microscopic analysis of the edges of
the YBCO was presented, the ion milling technique employed
necessarily leaves a rough surface
 on a scale of at least 1-5 nm, so that the tunneling
across each of these edges is most likely incoherent.  The thick
Au layer could modify the overall tunneling process, however.  The
much longer length scales $\lambda_J$ of the Josephson vortices,
which is only 1-2 orders of magnitude smaller than the edge
dimensions, allows the vortices to weakly interact, forming a
regular array. These experiments are suggestive of a dominant
$d$-wave OP.

However, we recall that Gim {\it et al.} studied YBCO/Pb SQUID's
prepared with a much wider selection of SQUID angles
$\theta$.\cite{Gim}  In those experiments, which did not include
$\theta=90^{\circ}$, the results were inconsistent with either a
predominant $d$- or $s$-wave OP component,\cite{Gim} but were
instead consistent with a $p$-wave polar state. Apparently, the
preparation of a $0$-junction or a $\pi$-junction depends upon the
ramping angle $\varphi$ of the bombarding ions, so that many pairs
of SQUID's with angles $\theta$ and $\theta+180^{\circ}$ contained
one 0-junction and one $\pi$-junction.\cite{Gim} Although the nice
zigzag experiments do demonstrate the reproducibility of the
0-junction and $\pi$-junction preparation at fixed $\varphi$ and
$\theta=90^{\circ}$ with YBCO/Nb junctions, those authors did not
report the results of   $\theta$ and $\varphi$ variation
studies.\cite{Smilde,HilgenkampNature}

We are presently engaged in calculating the single quasiparticle
and pair excitations at low $T$, both of which are contained in
the Green function matrix given  by Eqs. (\ref{gmnp}) and
(\ref{gmnm}). For brevity we have here only presented the
Josephson critical current across the grain boundaries, the
quasiparticle densities of states at the sites on the rows
adjacent to or near to the junction can be evaluated from the
imaginary part of the single particle $g^{11}_{nn}$ for
$n=0,1,\ldots$  Although the most commonly used procedure for
obtaining the zero-bias conductance peak on a surface is to use
the Bogoliubov-deGennes technique,\cite{TanakaKashiwaya} our Green
function technique with the appropriate surface boundary
conditions is equally valid,\cite{Barash,Kalenkov} but is
considerably more compact and powerful. With our Green function
technique, we are able to evaluate the critical current at
arbitrary grain boundary angles, and are not restricted to
specific interfaces such as $(pq0)|(p'q'0)$ for integral
$p,q,p',q'$. It is also much easier to generalize the tunneling
forms, although we have limited our treatment to the most commonly
studied specular and random cases. Although it would be rather
easy to generalize the tunneling matrix elements to forms
intermediate to these limiting cases for symmetric and straight
junctions,\cite{bks} for asymmetric junctions, such a
generalization is more difficult.  We have compared our results
with those obtained by Shirai {\it et al.} using a purely
numerical technique, and some of their numerical results at low
$T$ are in qualitative agreement with our results obtained from
Eq. (\ref{current}).  Our results with the nearly circular FS3 are
in fairly good agreement with those obtained with the GL model and
with microscopic calculations using a circular
FS.\cite{SigristRice,TKYBCO,Barash} The main drawback with the
Green function technique is the procedure for making the
calculations self-consistent. However, as noted by Tanaka and
Kashiwaya,\cite{TK1998} making the results self-consistent does
not usually result in qualitative changes. We are presently
engaged in  self-consistent calculations for various grain
boundaries of the $(pq0)|(p'q'0)$ types, as was done by Shirai
{\it et al.}\cite{Shirai} but we are using the near-neighbor and
next-nearest neighbor single quasiparticle matrix elements
appropriate for either the ARPES or FS2 FS's on Bi2212, and will
present our results elsewhere.\cite{ArnoldKlemmnext} We will also
present the quasiparticle densities of states for these more
realistic FS models elsewhere.\cite{ArnoldKlemmnext}

Finally, we remark that recent YBCO thin film structures have been
studied using $(001)|(103)$  thin film boundaries with
45$^{\circ}$ in-plane misalignment angles, and the results
interpreted using the Sigrist-Rice formula. \cite{Lindstrom,Bauch}
We have not yet analyzed such grain boundaries, because the Fermi
surface of YBCO is particularly complicated, and our FS2 and ARPES
FS's are likely to be inadequate to do so reliably.  Nevertheless,
from the results we obtained with these two FS's, it is evident
that the results obtained from such devices, the grain boundaries
of which have not been analyzed with atomic scale precision as was
done with the Bi2212 $c$-axis twist experiments,\cite{Klemmreview}
might also have a more conventional
explanation.\cite{Baselmans,Shaikhaidarov,Huang,Giazotto,Ketterson,Klapwijkreview}

It has been suggested that the experimental $I_c(\theta)$ observed
in $(001)$ tilt grain boundary junctions can only be understood in
terms of interface defects, regardless of the OP.\cite{Gurevich}
Of course, in Bi2212, it has long been established that a
significant amount of static defects exist not only on the top
cleaved surface,\cite{Lang} but throughout even the best single
crystals.\cite{Coppens,Moss,Grebille}
 Such defects combined with significant oxygen loss at the grain boundaries may be
  the most important source of the observed $\pi$-junctions in
the tricrystal and tetracrystal experiments, accounting for the
apparent inconsistency between those and  the $c$-axis bicrystal
and cross-whisker
experiments.\cite{Li,Takano,Latyshev,Klemmreview}

Finally, we recall that $\pi$-junctions are now routinely produced
using
 junctions of the SNS and SINIS varieties constructed from conventional
 superconductors,\cite{Baselmans,Shaikhaidarov,Huang,Giazotto,Ketterson,Klapwijkreview}
 so that a close reexamination of the tri- and
 tetracrystal experiments is warranted.  This is especially true
 in the case of the hole-doped cuprates.  We note that not only
 are $\pi$-junctions readily obtained with Nb, Al and their
 oxides,\cite{Baselmans} but the associated zero-bias conductance peaks and higher harmonic order Josephson phase relations in such
 conventional junctions have also been observed,\cite{Baselmans,Shaikhaidarov,Huang,Giazotto,Ketterson,Klapwijkreview}
 as in YBCO.\cite{Lindstrom,Bauch}

\section{Summary}

We  calculated the critical current  $I_c$ versus (001) in-plane
tilt angle $\theta$ for symmetric and asymmetric grain boundary
junctions, assuming the order parameter has either the $s$-,
extended-$s$-, or $d_{x^2-y^2}$-wave  form.  We used two Fermi
surfaces appropriate for hole-doped cuprates,  one appropriate for
electron-doped cuprates, and an extreme Fermi surface for
comparison purposes, and took account of the surface boundary
conditions appropriate for the interfaces. We studied both
tunneling limits of specular and random tunneling. An important
feature of our results is that the presence of the surface
boundary conditions allows for a finite $d$-wave critical current
across junctions in which the microscopic tunneling is random.
This is especially important because the tunneling across nearly
all
 in-plane (001) tilt junctions is expected to be
random on an atomic scale, even within each reconstructed facet.

 Although our results for electron-doped cuprates are in complete
agreement with the interpretation of the observation of the
tricrystal experiments on those compounds, our results for
hole-doped cuprates differ qualitatively with those obtained from
Ginzburg-Landau models,\cite{SigristRice,TKYBCO} and indicate that
the interpretation of the tricrystal experiments is complicated,
depending upon strong assumptions about the details of the Fermi
surface topology and the nature of the tunneling processes. In
addition, corresponding problems in  the interpretation of the
tetracrystal experiments on hole-doped cuprates persist, leading
to expectations of  an $I_c(\theta)$ inconsistent with many
experiments.\cite{Mannhartreview} We urge further in-plane $(001)$
tilt grain boundary experiments using films deposited by liquid
phase epitaxy to be made, and for high resolution transmission
electron microscopy studies of the pertinent grain boundaries to
be presented.\cite{Mannhartreview,Klemmreview,Eltsev}

\section*{Acknowledgments}
The authors would like to thank  T. Claeson, S. E. Shafranjuk, and
A. Yurgens for useful discussions.


\begin{thebibliography}{99}
\bibitem{TKreview}
C. C. Tsuei and J. R. Kirtley, Rev. Mod. Phys. {\bf 72}, 969
(2000).
\bibitem{Mannhartreview}
H. Hilgenkamp and J. Mannhart, Rev. Mod. Phys. {\bf 74}, 485
(2002).
\bibitem{Mueller}
K. A. M{\"u}ller, Phil. Mag. Lett. {\bf 82}, 279 (2002).
\bibitem{Klemmreview}
R. A. Klemm,  Phil. Mag. {\bf 85}, 801 (2005); Int. J. Mod. Phys.
B {\bf 12}, 2920 (1998).
\bibitem{pseudogap}
R. A. Klemm in {\it Nonequilibrium Physics at Short Times Scales.
Formation of Correlations}, edited by K. Morawetz (Springer,
Berlin, 2004), pp. 381-400.
\bibitem{ChaudhariLin}
P. Chaudhari and S. Y. Lin, Phys. Rev. Lett. {\bf 72}, 1084
(1994).
\bibitem{TKYBCO}
C. C. Tsuei, J. R. Kirtley, C. C. Chi, L. S. Yu-Jahnes, A. Gupta,
T. Shaw, J. Z. Sun, and M. B. Ketchen, Phys. Rev. Lett. {\bf 73},
593 (1994).
\bibitem{TKBi2212}
J. R. Kirtley, C. C. Tsuei, H. Raffy, Z. Z. Li, A. Gupta, J. Z.
Sun, and S. Megtert, Europhys. Lett. {\bf 36}, 707 (1996).
\bibitem{TKhole}
C. C. Tsuei, J. R. Kirtley, G. Hammer, J. Mannhart, H. Raffy, and
Z. Z. Li, Phys. Rev. Lett. {\bf 93}, 187004 (2004).
\bibitem{TKNCCO}
C. C. Tsuei and J. R. Kirtley, Phys. Rev. Lett. {\bf 85}, 182
(2000).
\bibitem{MannhartSQUID}
R. R. Schulz, B. Chesca, B. Goetz, C. W. Schneider, A. Schmall, H.
Bielefeldt, H. Hilgenkamp, J. Mannhart, and C. C. Tsuei, Appl.
Phys. Lett. {\bf 76}, 912 (2000).
\bibitem{Chesca}
B. Chesca, K. Ehrhardt, M. M{\"o\ss}le, R. Straub, D. Koelle, R.
Kleiner, and A. Tsukada, Phys. Rev. Lett. {\bf 90}, 057004 (2003).
\bibitem{Dynes}
A. G. Sun, D. A. Gajewski, M. B. Maple, and R. C. Dynes, Phys.
Rev. Lett. {\bf 72}, 2267 (1994); A. G. Sun, A. Truscott, A. S.
Katz, R. C. Dynes, B. W. Veal, and C. Gu, Phys. Rev. B {\bf 54},
6734 (1996).
\bibitem{Clarke}
R. Kleiner, A. S. Katz, A. G. Sun, R. Summer, D. A. Gajewski, S.
H. Han, S. I. Woods, E. Dantsker, B. Chen, K. Char, M. B. Maple,
R. C. Dynes, and J. Clarke, Phys. Rev. Lett. {\bf 76}, 2161
(1996).
\bibitem{Katz}
A. S. Katz, A. G. Sun, R. C. Dynes, and K. Char, Appl. Phys. Lett.
{\bf 66}, 105 (1995).

\bibitem{KleinerMoessle}
 M. M{\"o\ss}le and R. Kleiner, Phys. Rev. B {\bf 59}, 4486
 (1999).

\bibitem{Woods}
S. I. Woods, A. S. Katz, T. L. Kirk, M. C. de Andrade, M. B.
Maple, and R. C. Dynes, IEEE Trans. Appl. Supercond. {\bf 9}, 3917
(1999).

\bibitem{Li}
Qiang Li, Y. N. Tsay, M. Suenaga, R. A. Klemm, G. D. Gu, and N.
Koshizuka, Phys. Rev. Lett. {\bf 83}, 4160 (1999).

\bibitem{Takano}
Y. Takano, T. Hatano, S. Kawakami, M. Ohmi, S. Ikeda, M. Nagao, K.
Inomata, K. S. Yun, A. Ishii, A. Tanaka, T. Yamashita, and M.
Tachiki, Physica C {\bf 408-410}, 296 (2004).

\bibitem{Klemm}
R. A. Klemm, Phys. Rev. B {\bf 67}, 174509 (2003).

\bibitem{Latyshev}
Yu. I. Latyshev, A. P. Orlov, A. M. Nikitina, P. Monceau, and R.
A. Klemm, Phys. Rev. B {\bf 70}, 094517 (2004).

\bibitem{SigristRice}
M. Sigrist and T. M. Rice,  J.  Phys. Soc. Jpn. {\bf 61}, 4283
(1992).

\bibitem{TanakaKashiwaya}
Y. Tanaka and S. Kashiwaya, Phys. Rev. Lett. {\bf 74}, 3451
(1995); Phys. Rev. B {\bf 53}, R11957 (1996); {\it ibid.} {\bf
56}, 892 (1997).

\bibitem{Asano}
Y. Asano, Phys. Rev. B {\bf 64}, 014511 (2001); Y. Asano, Y.
Tanaka, and S. Kashiwaya, Phys. Rev. B {\bf 69}, 134501 (2004).

\bibitem{Barash}
Yu. S. Barash, H. Burkhardt, and D. Rainer, Phys. Rev. Lett. {\bf
77}, 4070 (1996).

\bibitem{Kalenkov}
M. S. Kalenkov, M. Fogelstr{\"o}m, and Yu. S. Barash, Phys. Rev. B
{\bf 70}, 184505 (2004).

\bibitem{Ilichev}
E. Il'ichev, M. Grajcar, R. Hlubina, R. P. J. IJsselsteijn, H. E.
Hoenig, H.-G. Meyer, A. Golubov, M. H. S. Amin, A. M. Zagoskin, A.
N. Omelyanchouk, and M. Yu. Kupriyanov, Phys. Rev. Lett. {\bf 86},
5369 (2001).


\bibitem{Shirai}
S. Shirai, H. Tsuchiura, Y. Asano, Y. Tanaka, J. Inoue, Y. Tanuma,
and S. Kashiwaya, J. Phys. Soc. Jpn. {\bf 72}, 2299 (2003). In
their Eq. (15), these authors defined their quasiparticle
dispersion  to be $\epsilon_{\bm q}=-(t+\zeta_1)\eta_{\bm
q}-\zeta_2\gamma_{\bm q}-(\mu+8Wn)$, with $\eta_{\bm q}=2(\cos
q_x+\cos q_y)$, their Eq. (17), in units of the lattice constant
$a=1$. Following their Eq. (21), the definitions $t=W$,
$\zeta_1=-0.19t$, $\zeta_2=0.0t$, $\mu=-0.2833t$, and the hole
density $n=\langle\sum_{\sigma}n_{{\bm r},\sigma}\rangle=0.85$
were given, where the summation is only over the two spin states.
 Putting these numbers into the above dispersion, one obtains
$\epsilon_{\bf q}=-1.606t(\cos q_x+\cos q_y+4.06)$, which never
vanishes, and hence does  not exhibit a Fermi surface. However,
the ${\bm q}$-independent part of this expression is inconsistent
with their Eq. (4) for the effective chemical potential,
$\tilde\mu=\mu+W\sum_{{\bm\rho},\sigma}\langle
n_{{\bm\rho},\sigma}\rangle$, where the summation is over the four
near-neighbor in-plane sites denoted by ${\bm\rho}$ and the two
spin states.  Hence,  the 8 in their Eq. (15) should be replaced
with the number 4 to avoid overcounting  the two spin states.
 With this modification, the correct dispersion of Shirai {\it et
al.} becomes $\epsilon_{\bf q}=-1.606t(\cos q_x+\cos q_y+1.94)$,
which has the Fermi surface at small pockets in the corners of the
first Brillouin zone shown in Fig. 2. Although Shirai {\it et al.}
may have intended to include this factor of 4 in their definition
of $n$,  which would have led to a Fermi surface resembling FS2 in
Fig. 2, from the equations in their manuscript  that they very
likely used the above dispersion corresponding in our notation to
$J_{||}=201$ meV, $\nu=0$, and $\mu=-1.94$, where $J_{||}$ can be
determined from their value of $\Delta=0.0799t$, which we set
equal to 10 meV.  However,  their strange results  for the
symmetric (mirror)  and   especially the straight (parallel)
18.4$^{\circ}$ junctions do not appear to correspond to any
results by other authors for any FS studied.

\bibitem{bks}
A. Bille, R. A. Klemm, and K. Scharnberg, Phys. Rev. B {\bf 64},
174507 (2001).

\bibitem{ArnoldKlemm}
G. B. Arnold and R. A. Klemm, Phys. Rev. B {\bf 62}, 661 (2000).

\bibitem{LiuKlemm}
S. H. Liu and R. A. Klemm, Phys. Rev. Lett. {\bf 73}, 1019 (1994).

\bibitem{Cai}
X. Y. Cai, A. Gurevich, I.-F. Tsu, D. L. Kaiser, S. E. Babcock,
and D. C. Larbalestier, Phys. Rev. B {\bf 57}, 10951 (1998).

\bibitem{Chisholm} M. F. Chisholm and S. J. Pennycook, Nature
(London) {\bf 351}, 47 (1991).

\bibitem{YanHu}
X.-Z. Yan and C.-R. Hu, Phys. Rev. Lett. {\bf 83}, 1656 (1999).

\bibitem{ArnoldKlemmnext}
G. B. Arnold and R. A. Klemm, unpublished.



\bibitem{TK1998}
Y. Tanaka and S. Kashiwaya, Phys. Rev. B {\bf 58}, R2948 (1998).
\bibitem{kars} R. A. Klemm, G. Arnold, C. T. Rieck, and K. Scharnberg, Phys. Rev. B {\bf 58},
14203 (1998).

\bibitem{Eltsev}
Yu. Eltsev, K. Nakao, Y. Yamada, I. Hirabayashi, Y. Ishimaru, K.
Tanabe, Y. Enomoto, J. Wen, and N. Koshizuka, Physica C {\bf 367},
24 (2002).

\bibitem{AB}
V. Ambegaokar and A. Baratoff, Phys. Rev. Lett. {\bf 10}, 486
(1963); {\it ibid.} {\bf 11}, 104 (1963).

\bibitem{Ishimaru}
Y. Ishimaru, J. Wen, N. Koshizuka, and Y. Enomot, Phys. Rev. B
{\bf 55}, 11851 (1997).

\bibitem{Miller}
X. F. Zhang, D. J.  Miller and J. Talvacchio,   J. Mater. Res.
{\bf 11}, 2440 (1996).

\bibitem{Gurevich}
A. Gurevich and E. A. Pashitskii, Phys. Rev. B {\bf 57}, 13878
(1998) and references therein; {\it ibid.} {\bf 63}, 139901(E)
(2001).

\bibitem{Lang}
K. M. Lang, V. Madhavan, J. E. Hoffman, E. W. Hudson, H. Eisaki,
J. Z. Sun, and S. Megtert, Nature (London) {\bf 415}, 412 (2002).

\bibitem{Coppens} Y. Gao, P. Coppens, D. E. Cox, and A. R. Moodenbaugh,
 Acta Cryst. A {\bf 49}, 141 (1993); P. Coppens, P. Lee, G. Yang, and H. S. Sheu,
  J. Phys. Chem. Solids {\bf 52}, 1267 (1991); V. Petricek, Y. Gao, P. Lee,
  and P. Coppens,
Phys. Rev. B {\bf 42}, 387 (1990).

\bibitem{Moss}
X. B. Kan and S. C. Moss, Acta Cryst. B {\bf 48}, 122 (1992).

\bibitem{Grebille}
D. Grebille, H. Leligny, A. Ruyter, P. Labb{\'e}, and D. Raveau,
Acta Cryst. B {\bf 52}, 628 (1996).



\bibitem{Smilde} H. J. H. Smilde, Ariando, D. H. A. Blank, G. J.
Gerritsma, H. Hilgenkamp, and H. Rogalla,  Phys. Rev. Lett. {\bf
88}, 057004 (2002).

\bibitem{HilgenkampNature}
H. Hilgenkamp, Ariando, H.-J. H. Smilde, D. H. A. Blank, G.
Rijnders, H. Rogalla, J. R. Kirtley, and C. C. Tsuei, Nature
(London) {\bf 422}, 50 (2003).

\bibitem{Mathai}
A. Mathai, Y. Gim, R. C. Black, A. Amar, and F. C. Wellstood,
Phys. Rev. Lett. {\bf 74}, 4523 (1995); Anna Mathai, Ph. D.
Thesis, University of Maryland, 1995.

\bibitem{Gim}
Y. Gim, A. Mathai, R. C. Black, A. Amar, and F. C. Wellstood, J.
Phys. (Paris) I {\bf 6}, 2299 (1996);  Yonggyu Gim, Ph. D. Thesis,
University of Maryland, 1996.

\bibitem{Baselmans}
J. J. A. Baselmans, A. F. Morpurgo, B. J. van Wees, and T. M.
Klapwijk, Nature (London) {\bf 397}, 43 (1999); J. J. A.
Baselmans, B. J. van Wees, and T. M. Klapwijk, Appl. Phys. Lett.
{\bf 79}, 2940 (2001); J. J. A. Baselmans, T. T. Heikkil{\"a}, B.
J. van Wees, and T. M. Klapwijk, Phys. Rev. Lett. {\bf 89}, 207002
(2002).

\bibitem{Shaikhaidarov}
R. Shaikhaidarov, A. F. Volkov, H. Takayanagi, V. T. Petrashov,
and P. Delsing, Phys. Rev. B {\bf 62}, R14649 (2000).

\bibitem{Huang}
J. Huang, F. Pierre, T. T. Heikkil{\"a}, F. K. Wilhelm, and N. O.
Birge, Phys. Rev. B {\bf 66}, 020507(R) (2002).

\bibitem{Giazotto}
F. Giazotto, T. T. Heikkil{\"a}, F. Taddei, R. Fazio, J. P.
Pekola, and F. Beltram, Phys. Rev. Lett. {\bf 92}, 137001 (2004).

\bibitem{Ketterson}
I. P. Nevirkovets, S. E. Shafranjuk, and J. B. Ketterson, Phys.
Rev. B {\bf 68},   024514 (2003).

\bibitem{Klapwijkreview}
T. M. Klapwijk, J. Supercond.:  Incorp. Novel Magn. {\bf 17}, 593
(2004).

\bibitem{Lindstrom}
T. Lindstr{\"o}m, S. A. Charlebois,  A. Ya. Tzalenchuk, Z. Ivanov,
M. H. S. Amin, and A. M. Zagoskin, Phys. Rev. Lett. {\bf 90},
117002 (2003).

\bibitem{Bauch}
T. Bauch, F. Lombardi, F. Tafuri, A. Barone, G. Rotoli, P.
Delsing, and T. Claeson, Phys. Rev. Lett. {\bf 94}, 087003 (2005).
\end{thebibliography}
\end{document}